\documentclass[
superscriptaddress,
amsmath,
amssymb,
aps,
prb,
twocolumn,
longbibliography,
]{revtex4-2}

\usepackage{color}
\usepackage[T1]{fontenc}  
\usepackage{graphicx}
\usepackage{dcolumn}
\usepackage{bm}
\usepackage[colorlinks=true, linkcolor=blue, citecolor=blue, urlcolor=blue]{hyperref}
\usepackage{placeins} 
\usepackage{float} 
\usepackage{cleveref} 

\newcommand*\Laplace{\mathop{}\!\mathbin\bigtriangleup}

\begin{document}

\preprint{APS/123-QED} 

\title{Experimental study of Su-Schrieffer-Heeger edge modes for water waves \\ in linear and nonlinear regime}

\author{Adam Anglart} \email{adam.anglart@espci.fr} \affiliation{PMMH, ESPCI Paris, Université PSL, CNRS, \\ Sorbonne Université, Université Paris Cité, Paris, France}
\author{Paweł Obrępalski} \affiliation{PMMH, ESPCI Paris, Université PSL, CNRS, \\ Sorbonne Université, Université Paris Cité, Paris, France}
\author{Agnès Maurel} \affiliation{Institut Langevin, ESPCI Paris, Université PSL, CNRS, Paris, France}
\author{Philippe Petitjeans} \affiliation{PMMH, ESPCI Paris, Université PSL, CNRS, \\ Sorbonne Université, Université Paris Cité, Paris, France}
\author{Vincent Pagneux} \affiliation{LAUM, Le Mans Université, CNRS, Le Mans, France}

\date{\today}

\begin{abstract}
This paper experimentally investigates topologically protected edge modes in a water wave channel through a direct geometric mapping to the one-dimensional Su-Schrieffer-Heeger (SSH) model. By designing a periodic channel with alternating widths, we replicate the key features of the SSH model, leading to the emergence of robust zero-energy sloshing edge modes localized at the boundaries. Experimental data show excellent agreement with theoretical predictions, supported by two-dimensional numerical simulations. In the nonlinear regime, two distinct bifurcations are observed, indicating the appearance of secondary resonances. This study highlights the relevance of the SSH model for water wave systems and provides an accessible method to explore topological edge states in classical wave systems.
\end{abstract}

\maketitle
\section{Introduction}
Metamaterials are typically defined as artificially structured materials designed to control the propagation of waves in unconventional ways. They are composed of unit cells whose dimensions are much smaller than the incident wavelength. This subwavelength condition is crucial for the homogenization process, which plays an essential role in deriving the effective medium approximation \cite{oleinik2009mathematical}. However, unique wave phenomena can also emerge in structures where the wavelength is comparable to or exceeds the unit cell size \cite{rybin2015phase, yi2016role}. The study of such phenomena has been explored in the context of topological insulators, whose investigation began after the discovery of the quantum Hall effect \cite{thouless1982quantized, prange1990quantum} and has since been extended to various wave systems, including acoustics \cite{Zhao2018Topological, Zhang2018Topological, Xue2019Realization, Ma2019Topological, coutant2021acoustic}, photonics \cite{Khanikaev2013Photonic, Lu2014Topological, Khanikaev2017Two, Ozawa2019Topological}, phononics \cite{Susstunk2015Observation, Brendel2018Snowflake, SerraGarcia2018Observation}, and mechanics \cite{Edwards1967Statistical, Nash2015Topological, Huber2016Topological, Chen2016Topological}. 

A key feature of topological metamaterials is the existence of edge modes - topologically protected states localized at boundaries \cite{bernevig2013topological}. These modes are robust against perturbations that preserve the underlying symmetries of the system. In bipartite systems, the stability of edge modes is ensured by chiral symmetry, which protects them even when disorder randomizes coupling strengths (as long as the bipartite structure and symmetry are preserved) \cite{mudry2003density, ramachandran2017chiral,coutant2021acoustic}. These results highlight the broader resilience of topological edge modes under perturbations that respect the symmetry of the system.

One of the most well-known frameworks for describing topologically protected states is the Su-Schrieffer-Heeger model (SSH) \cite{ssh11, ssh21}, which has been studied, e.g., in acoustics \cite{li2018schrieffer, esmann2018topological, yan2020acoustic,A3_coutant2022subwavelength}. In the context of water waves, it has been studied in an array of water tanks connected by narrow channels \cite{yang2016topological} and in systems featuring edge states over a structured bottom \cite{B1wu2018topological, B2guan2024fluctuation}. In most of the cases, the theoretical approach relies on coupled resonator systems within tight-binding approximation (TBA) and band inversion coming with gap closing.

    \begin{figure}[b]
    \centering
    \includegraphics[width=.49\textwidth]{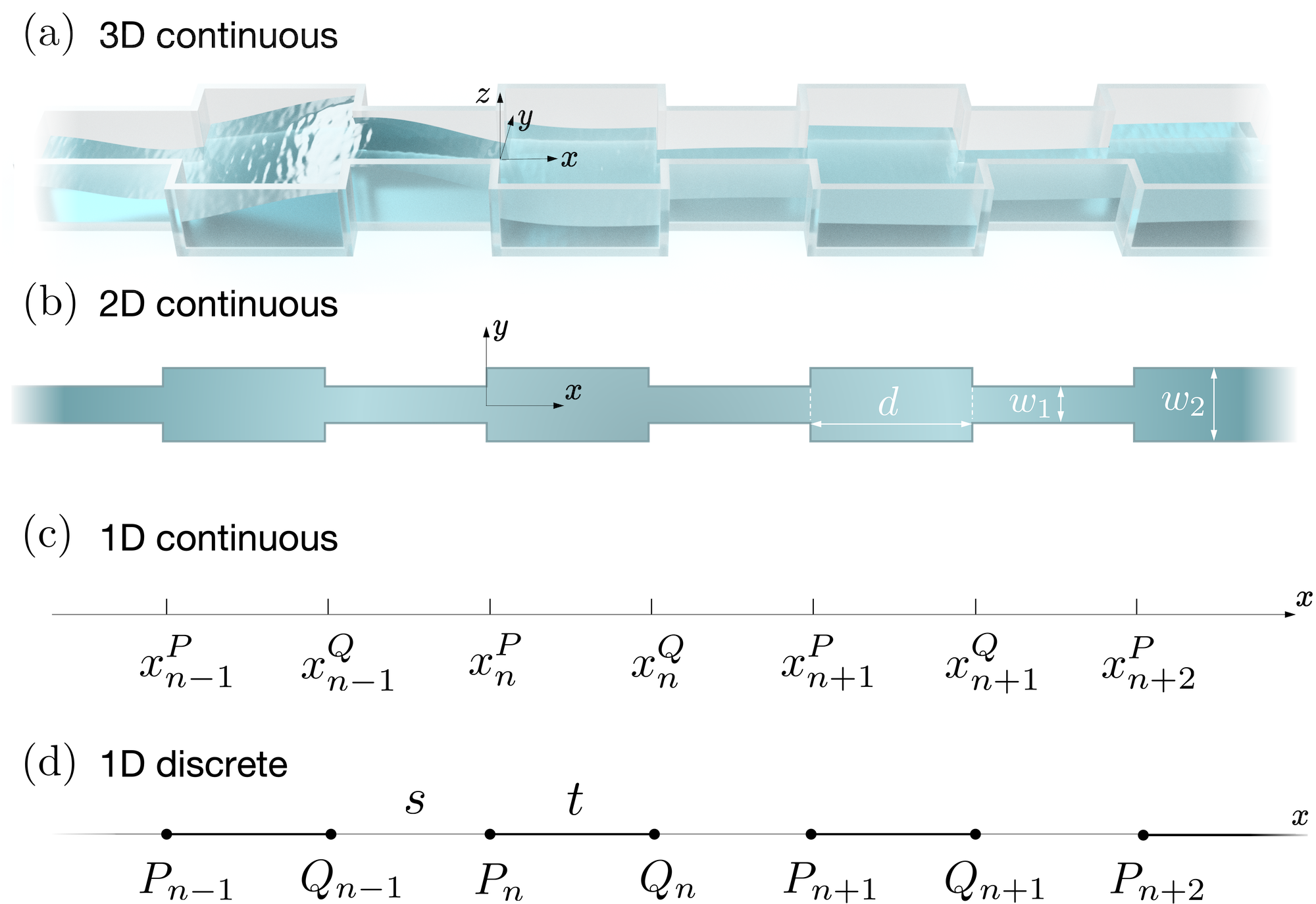}
    \caption{(a) Scheme of the fully three-dimensional water wave problem. (b) Scheme of the two-dimensional periodic channel consisting of the cells with different widths $w_1$, $w_2$ and the length $d$, with water waves following Eq.~\eqref{helmSSH}. (c) One-dimensional continuous approximation (Eq.~\eqref{1dcont}) leaves us with $\eta$ depending only on $x$. On the axis, we identify points corresponding to the change of the channel widths, where Eq.~\eqref{jump1} applies. (d) Scheme of the one-dimensional discrete SSH model (Eqs.~\eqref{ssh1u} and \eqref{ssh2u}) corresponding to the water wave channel.}
    \label{schemeSSH}
    \end{figure} 

In this experimental work, we use a different approach that is not using TBA with resonators. The periodic geometry of the water-wave channel allows to straightforwardly implement the SSH model, enabling us to observe topologically protected edge modes in the laboratory-scale setup. This study also examines sloshing modes in both linear and nonlinear regimes, highlighting how edge modes persist under nonlinear conditions, similar to the persistence of topological edge breathers in nonlinear SSH lattices \cite{B3johansson2023topological}. Additionally, the role of boundary conditions in shaping the edge states is investigated, providing a more comprehensive understanding of topological phenomena in water-wave systems. This setup offers a practical way to study the connection between topology, wave behavior, and nonlinearity in water-wave systems.

 \section{Model reduction from 2D Helmholtz to 1D SSH}

    \begin{figure}[b]
    \centering
    \includegraphics[width=0.485\textwidth]{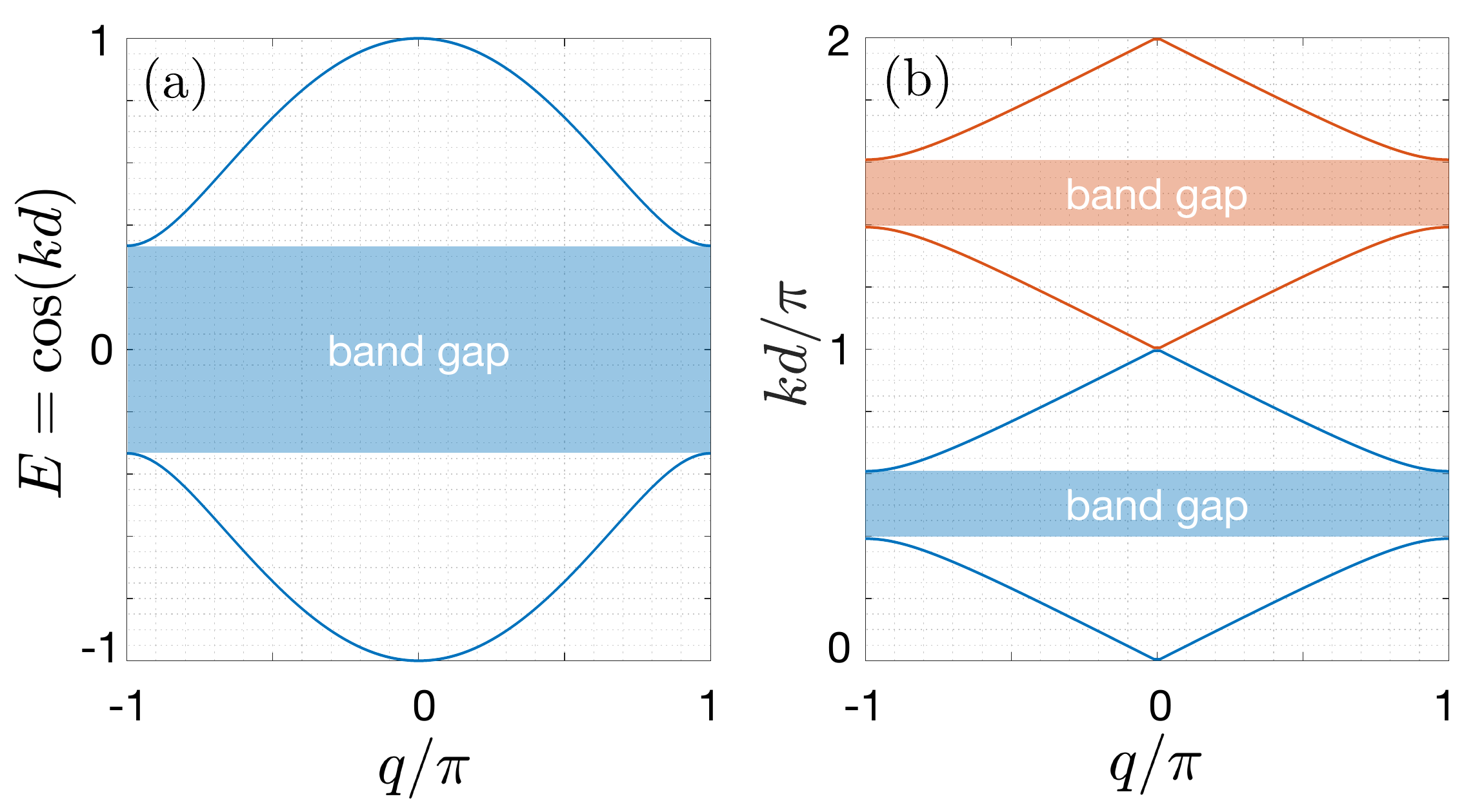}
    \caption{Dispersion relation of the SSH model \eqref{dispersionSSH}. (a) Pseudoenergy $E$ as a function of the Bloch wavenumber $q$ for $s=1/3$. (b) Undimensionalized wavenumber $k$ over the Bloch wavenumber $q$ for $s=1/3$.}
    \label{fig:ssh1}
    \end{figure} 

In this section we follow the same lines as in \cite{coutant2021acoustic,A3_coutant2022subwavelength} where, for a waveguide with connected sements of equal length, the initially 3D continuous problem is transformed into a 1D continuous approximation and then to a 1D discrete problem. Note that the same approach can be applied to network of connected segments to obtain a 2D discrete problem as well \cite{A4_zheng2019observation,A5_coutant2020robustness,A7_coutant2021topological,A6_coutant2024topologically}. Here, a water wave channel with a constant depth $h$ and vertical walls, characterized by segments of periodically alternating widths $w_1$, $w_2$, but equal lengths $d$, is considered, as shown in Fig.~\ref{schemeSSH}(a). Neglecting the effect of viscosity and in the linear regime, the free surface elevation $\eta(x,y)$, with time harmonic convention $\mathrm{e}^{-\mathrm{i}\omega t}$, satisfies the two-dimensional Helmholtz equation with the homogenous Neumann boundary condition corresponding to vanishing normal velocity on the walls \cite{A1_mei2005theory,A2_chabchoub2016time}

    \begin{align}
    \Laplace{\eta} +k^2 \eta =0, \nonumber \\
    \mathbf{n}   \cdot \nabla \eta = 0  \text{ on walls},
    \label{helmSSH}
    \end{align}
where $\mathbf{n}$ is a vector normal to the boundaries of the channel and the wavenumber $k$ satisfies the dispersion relation for water waves $\omega^2=gk\tanh(kh)$, where $g$ denotes the gravitational acceleration, $\omega$ is the frequency, and $h$ is the depth of the channel. Then, we further simplify the 2D Helmholtz equation by using the monomode approximation assuming sufficiently large wavelengths (low frequencies; with the cutoff $k<\pi/w_2$). This allows to transition to a one-dimensional, continuous wave equation
    \begin{align}
    \eta'' + k^2 \eta=0,
    \label{1dcont}
    \end{align}
where now $\eta=\eta(x)$ depends only on the horizontal direction $x$. At each cross section, corresponding to changes in channel width, we ensure the continuity of the free surface elevation and its derivative (representing flow rate) using the following jump conditions
    \begin{align}
    \left[\eta\right]=0 \quad \text{and} \quad \left[w\eta'\right]=0, \label{jump1}
    \end{align}
where $[f]=f^+-f^-$ describes the difference at each cross section from the right side limit ($f^+$) and the left side limit ($f^-$). After reducing the two-dimensional continuous model to its one-dimensional counterpart, we can further reduce the description to a one-dimensional discrete framework. The channel features two types of cross sections, designated as $P$ and $Q$. Cross section $P$ corresponds to transitions from $w_1$ to $w_2$, while $Q$ represents the transition from $w_2$ to $w_1$. Leveraging the known analytical solution to the equation \eqref{1dcont} under condition \eqref{jump1}, we derive the relationship between the neighboring points (here for sections $Q$)
    \begin{align}
    \eta(x_n^Q)&=\eta(x_n^P) \cos{kd} + \frac{\eta'(x_n^{P+})}{k}\sin{kd}, \label{eqssh1}\\
    \eta(x_{n-1}^Q)&=\eta(x_n^P) \cos{kd} - \frac{\eta'(x_n^{P-})}{k}\sin{kd}. \label{eqssh2}
    \end{align}
By multiplying equation \eqref{eqssh1} by $w_2$, equation \eqref{eqssh2} by $w_1$, subsequently summing the results, and taking advantage of \eqref{jump1} we obtain
    \begin{align}
    w_2\eta(x_n^Q) + w_1\eta(x_{n-1}^Q) = (w_1+w_2)\eta(x_n^P) \cos{kd}.
    \end{align}
Applying a similar methodology to sections $P$, we derive
    \begin{align}
    w_2\eta(x_n^P) + w_1\eta(x_{n+1}^P) = (w_1+w_2)\eta(x_n^Q) \cos{kd}.
    \end{align}
The coupling coefficients, $s$ and $t$, are introduced as
    \begin{align}
    s=\frac{w_1}{w_1+w_2} \quad \text{and} \quad t=\frac{w_2}{w_1+w_2}.
    \label{coupling}
    \end{align}
These coefficients are solely dependent on the geometry of the system, are positive, and collectively sum to unity ($s+t=1$). Identifying 
    \begin{align}
    E\equiv\cos{kd},
    \end{align}
and for the sake of simplicity defining
    \begin{align}
    Q_n \equiv  \eta(x_n^Q) \quad \text{and} \quad P_n\equiv \eta(x_n^P),
    \end{align}
we can ultimately describe our system by
    \begin{align}
    sQ_{n-1}+tQ_n&=EP_n, \label{ssh1u}\\
    tP_n+sP_{n+1}&=EQ_n. \label{ssh2u}
    \end{align}
which effectively represents an eigenvalue problem
    \begin{align}
    \mathbb{H}\mathbf{X}=E\mathbf{X},
    \end{align}
where
    \begin{align}
    \mathbb{H}=\begin{pmatrix}
      \ddots & \ddots &  &    &    &      \\
      \ddots & 0      & s &   &    &      \\
            & s      & 0 & t &   &       \\
            &        & t & 0 & s &       \\
            &        &    & s & 0 & \ddots  \\
            &        &    &  &  \ddots & \ddots  \\
    \end{pmatrix} \quad \text{and} \quad \mathbf{X}=\begin{pmatrix}
      \vdots \\
      Q_{n-1} \\
      P_n \\
      Q_n \\
      P_{n+1} \\
      \vdots \\
    \end{pmatrix}.
    \end{align}
The above representation of our system \textbf{precisely aligns with the SSH model} \cite{ssh21}. Given that the SSH Hamiltonian $\mathbb{H}$ is solely geometry-dependent, the eigenfrequencies $k$ are directly deduced from the eigenvalues of $\mathbb{H}$. Furthermore, the pseudoenergy $E(k)=\cos{kd}$ closely mirrors its counterpart in the SSH model.

To determine the dispersion relation of the system, we adopt the Bloch wave solution
    \begin{align}
    P_n=P\mathrm{e}^{\mathrm{i}qn} \quad \text{and} \quad Q_n=Q\mathrm{e}^{\mathrm{i}qn},
    \label{bloch}
    \end{align}
where $q$ is the Bloch wavenumber. Substituting \eqref{bloch} into \eqref{ssh1u} and \eqref{ssh2u} yields the following eigenvalue problem 
    \begin{align}
    \begin{pmatrix}
        0 & s\mathrm{e}^{-\mathrm{i}q} + t \\
        s\mathrm{e}^{\mathrm{i}q} + t & 0
    \end{pmatrix}
    \begin{pmatrix}
      P\\
      Q
    \end{pmatrix} = E \begin{pmatrix}
      P\\
      Q
    \end{pmatrix}.
    \end{align}
This $2\times2$ Hamiltonian matrix, characteristic of the periodic one-dimensional SSH system, facilitates the direct determination of the dispersion relation of the system
    \begin{align}
    E=\cos{kd} =\pm \sqrt{s^2 + 2st\cos q +t^2}.
    \label{dispersionSSH}
    \end{align}
Due to the chiral symmetry of the system \cite{ssh21}, the dispersion relation exhibits symmetry around $E=0$ [Fig.~\ref{fig:ssh1}(a)] that can be unwound as $kd=(m+1/2)\pi$ ($m=0,1,2\dots$) [Fig.~\ref{fig:ssh1}(b)]. Furthermore, the zero-energy mode ($E=0$) emerges when the wavelength is quadruple the length of segment $d$. This relation can be straightforwardly derived as $E=0 \rightarrow kd=\pi/2 \rightarrow \lambda=4d$.

    \begin{figure}[t]
    \centering
    \includegraphics[width=0.485\textwidth]{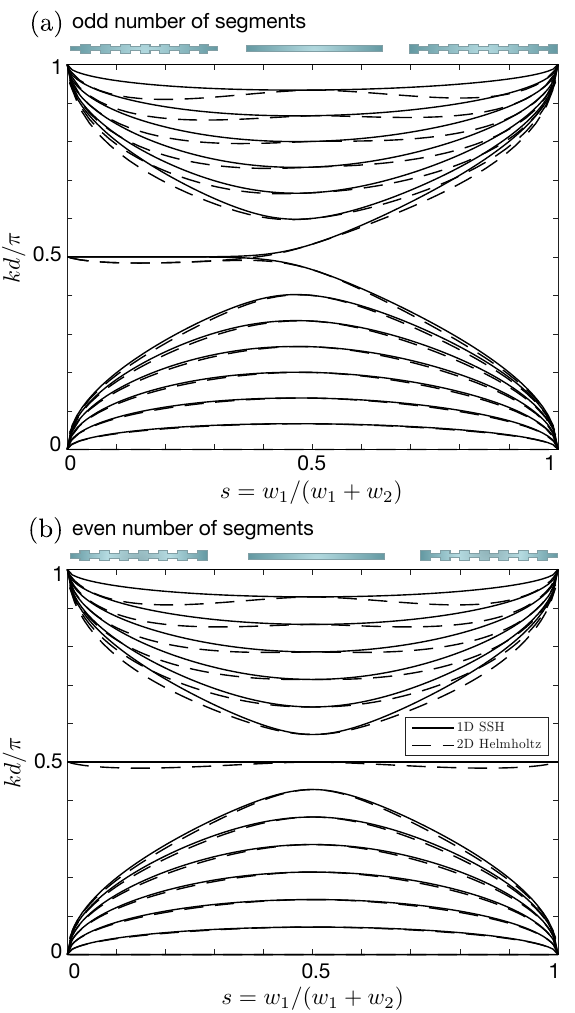}
    \caption{Eigenvalues of the channel with (b) an odd number of segments ($2N+1=15$) and (b) an even number of segments ($2N=14$) obtained by the SSH model (plain curves) and the numerical simulation (dashed curves) as a function of the parameter $s$.}
    \label{ssh2}
    \end{figure} 

    \begin{figure*}[t]
    \centering
    \includegraphics[width=\textwidth]{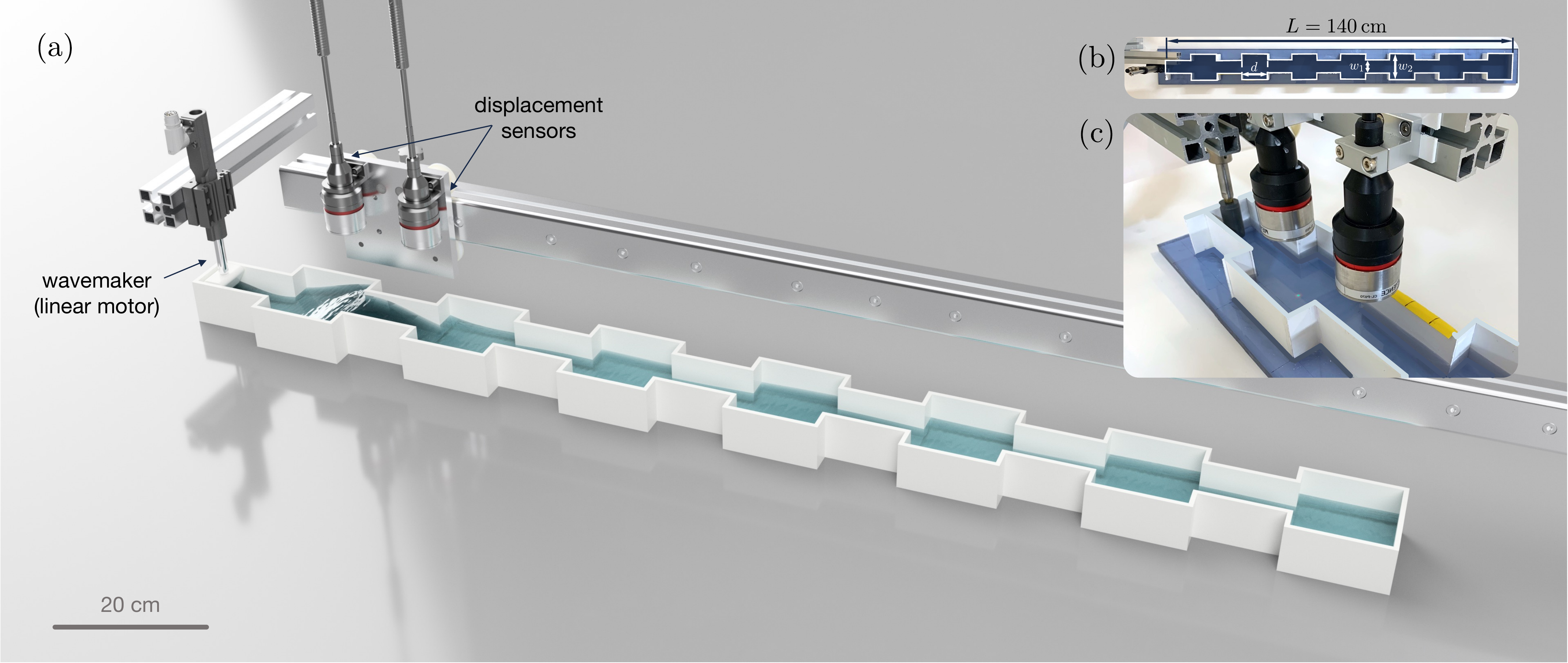}
    \caption{(a) Conceptual view of the channel, including the wavemaker placed at the left end of the channel, and two confocal displacement sensors attached to a movable trolley. (b) Photograph of the periodic channel used in the experiments, showing a top view with the wavemaker positioned at the left end. (c) Photograph of the left end of the channel with the wavemaker and the two confocal displacement sensors suspended above the channel.}
    \label{setup}
    \end{figure*} 

To demonstrate the applicability of the SSH model for water waves, we analyze a finite symmetric channel comprising an odd number of segments, 2N + 1 = 15, alongside its asymmetric counterpart with an even number of segments, 2N = 14.  The channel is closed on both ends, imposing a homogenous Neumann boundary condition at each wall, ensuring that the normal velocity vanishes. This setup corresponds to applying the model within a cavity. Consequently, the Hamiltonian and the vector of variables in the eigenvalue problem $\mathbb{H}\mathbf{X}=E\mathbf{X}$ are written as follows
    \begin{align}
    \mathbb{H} =
    \begin{pmatrix}
        0   & 1   & 0   & \dots & \dots & \dots & 0 \\
        s   & 0   & t   &       &       &       & \vdots \\
        0   & t   & 0   & s     &       &       & \vdots \\
        \vdots &   & \ddots & \ddots & \ddots &   & \vdots \\
        \vdots &   &       & t     & 0     & s   & 0 \\
        \vdots &   &       &       & s     & 0   & t \\
        0   & \dots & \dots & \dots & \dots & 1   & 0 \\
    \end{pmatrix} \quad \text{and} \quad \mathbf{X}=\begin{pmatrix}
        Q_0 \\
        P_1 \\
        Q_1 \\
        \vdots\\
        Q_{N-1} \\
        P_N \\
        Q_N\\
    \end{pmatrix}.
    \label{even}
    \end{align}
Eigenvalues derived from the SSH model for odd-segment scenarios are depicted in Fig.~\ref{ssh2}(a), while those for even-segment cases are presented in Fig.~\ref{ssh2}(b) (for all possible values of $s$ and $t=1-s$). In the odd case (\ref{ssh1u}, \ref{ssh2u}) it is straightforward to obtain an explicit form of the zero-energy mode localized at the edge for the system \eqref{even}
    \begin{align}
      \begin{pmatrix}
        P_n \\
        Q_n
      \end{pmatrix}= c \begin{pmatrix}
        0 \\
        1 
      \end{pmatrix} \left(-\frac{s}{t}\right)^n,
      \label{edgewave}
    \end{align}
where $c$ is the normalization constant.

\section{Numerical simulations}
To evaluate the applicability of the SSH model that is a 1D approximation, we numerically solve the full two-dimensional eigenvalue problem for the even and odd cases separately
    \begin{align}
    \Laplace{\eta} = -k^2 \eta, \nonumber \\
    \mathbf{n}   \cdot \nabla \eta = 0  \text{ on walls}.
    \label{helmSSHeigen}
    \end{align}
The solution is derived using the Finite Element Method, setting parameters at $w_1=0.05$, $w_2=0.1$, $d=0.1$, $L=1.4$ ($L$ being the total length of the channel), with results displayed in Fig.~\ref{ssh2}(a) for the scenario involving an odd number of segments, and in Fig.~\ref{ssh2}(b) for the configuration with an even number of segments. It is evident that, in the case of an even number of segments [$2N=14$, as shown in Fig.~\ref{ssh2}(b)], the system accommodates a single localized mode of zero energy. For $s < t$ the edge wave localizes on the left side of the channel, and \textit{vice versa}, for $s > t$ the mode localizes on the right side. This behavior is readily explicable upon reexamining the formulation of the edge wave \eqref{edgewave} as derived from the SSH model. It appears that only the edge segments where $w_1<w_2$ are able to host the edge wave. 
 
    \begin{figure*}[t]
    \centering
    \includegraphics[width=\textwidth]{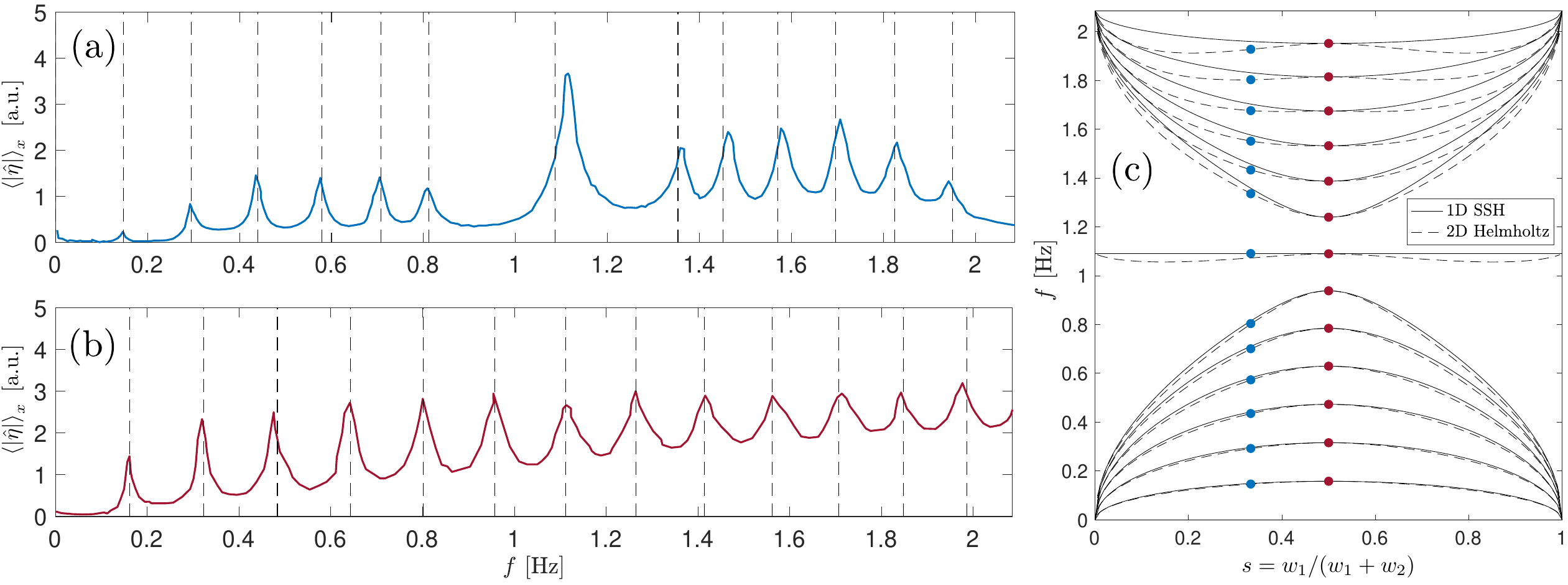}
    \caption{Averaged spectrum of the free-surface elevation obtained by averaging the Fourier spectra measured at all sensor positions (spaced every 2.5 cm along $x$-direction) (a) for $s=1/3$ (blue curve) with a 2D numerical prediction (dashed lines) and (b) for $s=1/2$ (orange curves) with a numerical prediction (dashed lines). (c) Comparison of the 1D (plain lines) and the 2D (dashed lines) prediction of the eigenvalues with the experimental values for $s=1/3$ (blue dots) and $s=1/2$ (orange dots).}
    \label{exp1}
    \end{figure*}

 Analyzing the case where we have an odd number of segments [$2N+1=15$, Fig.~\ref{ssh2}(a)], we see that for $s<t$, our system can host two zero energy modes - one localized at the left end and the other localized at the right end. For $s>t$, the edge mode disappears, as our system is unable to host the localized edge wave since on both ends we have segments with $w_1>w_2$. It is worth mentioning that for $s=t$, we obtain the case where the channel is rectangular, thus the localization of the zero energy mode does not occur. The discrepancies between the eigenvalues derived from the SSH model and those from the numerical solution can be explained by the fact that one-dimensional approximation is not able to account for near field effects close to each cross section changes \cite{coutant2021acoustic}. Nonetheless, the one-dimensional approximation remains satisfactorily precise, provided that the aspect ratio of the segment remains sufficiently small and the wavelength exceeds the width of the channel. 

\section{Experimental results}
\subsection{Experimental setup}
The experimental setup comprises a periodic channel [illustrated in Fig.~\ref{setup}(a)] with a length of $L=140 \, \mathrm{cm}$, featuring alternating widths of $w_1= 5 \, \mathrm{cm}$ and $w_2=10 \, \mathrm{cm}$, corresponding to a parameter setting of $s=1-t=1/3$. In order to facilitate the measurement of the topological edge mode, we choose the case of one isolated edge mode (even number of segments); thus, the channel consists of 14 segments [as in Fig.~\ref{ssh2}(b)] of length $d=10 \, \mathrm{cm}$ with a constant water depth of $h=2 \, \mathrm{cm}$. Note that we have selected the number of segments large enough in order to see the whole exponential decay of the edge wave \eqref{edgewave} (the edge mode has a negligible amplitude on the other end). The aspect ratio of the cells is small enough to apply the 1D approximation leading to the discrete SSH model, as shown in the previous section. On the other hand, from the water waves perspective, the aspect ratio has to be sufficiently big in order to avoid the detrimental effects of the meniscus that forms on the walls of the channel and whose size is of the order of $2 \, \text{mm}$. Indeed, in channels with small aspect ratios, the presence of the meniscus could induce undesirable shifts in the eigenfrequencies \cite{pawel, eduardo,monsalve2022space}. Additionally, as a reference, a rectangular channel with identical length and a constant width of $w=5 \, \mathrm{cm}$ was constructed in order to verify the regular modes of a straight cavity ($s=t$). The realization of the point source is done by placing the wavemaker consisting of the linear motor with cylindrical tip [Fig.~\ref{setup}(b)] of the diameter $\phi_s = 2 \, \mathrm{cm}$ and that is placed on the left end of the channel in the segment of smaller width, 0.5 cm from the wall. The wavemaker realizes the vertical motion corresponding to a chirp signal, whose length is $t=60 \, \mathrm{s}$ and whose frequency spectrum varies from 0.1 to 2.5 Hz (the cutoff frequency corresponding to $k = \pi/w_2$). The amplitude of the wavemaker motion ranges from $A_s = 0.5 \, \mathrm{mm}$ to $A_s = 15 \, \mathrm{mm}$, allowing us to cover both linear and nonlinear regimes for the water wave behavior. 

\begin{figure}[b]
    \centering
    \includegraphics[width=0.485\textwidth]{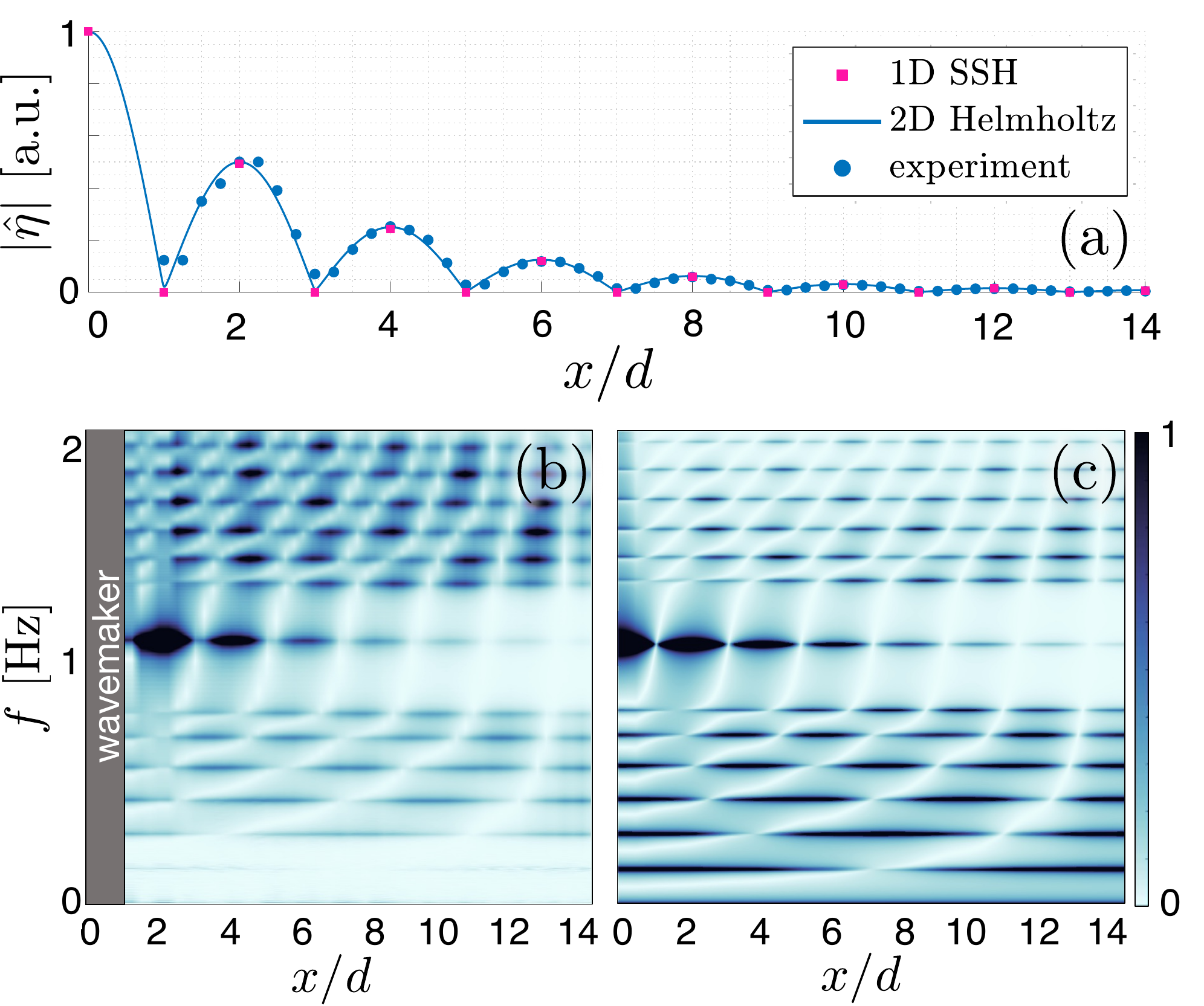}
    \caption{(a) Edge mode localized at the left side of the channel obtained experimentally (blue dots), by 2D simulation (averaged along $y$, plain curve), and by the SSH model \eqref{edgewave} (pink symbols). (b) Experimental results for $s=1/3$. (c) Absolute value of the free surface elevation obtained using Finite Element Method.}
    \label{exp2}
    \end{figure} 

    \begin{figure*}[]
        \centering
        \includegraphics[width=\textwidth]{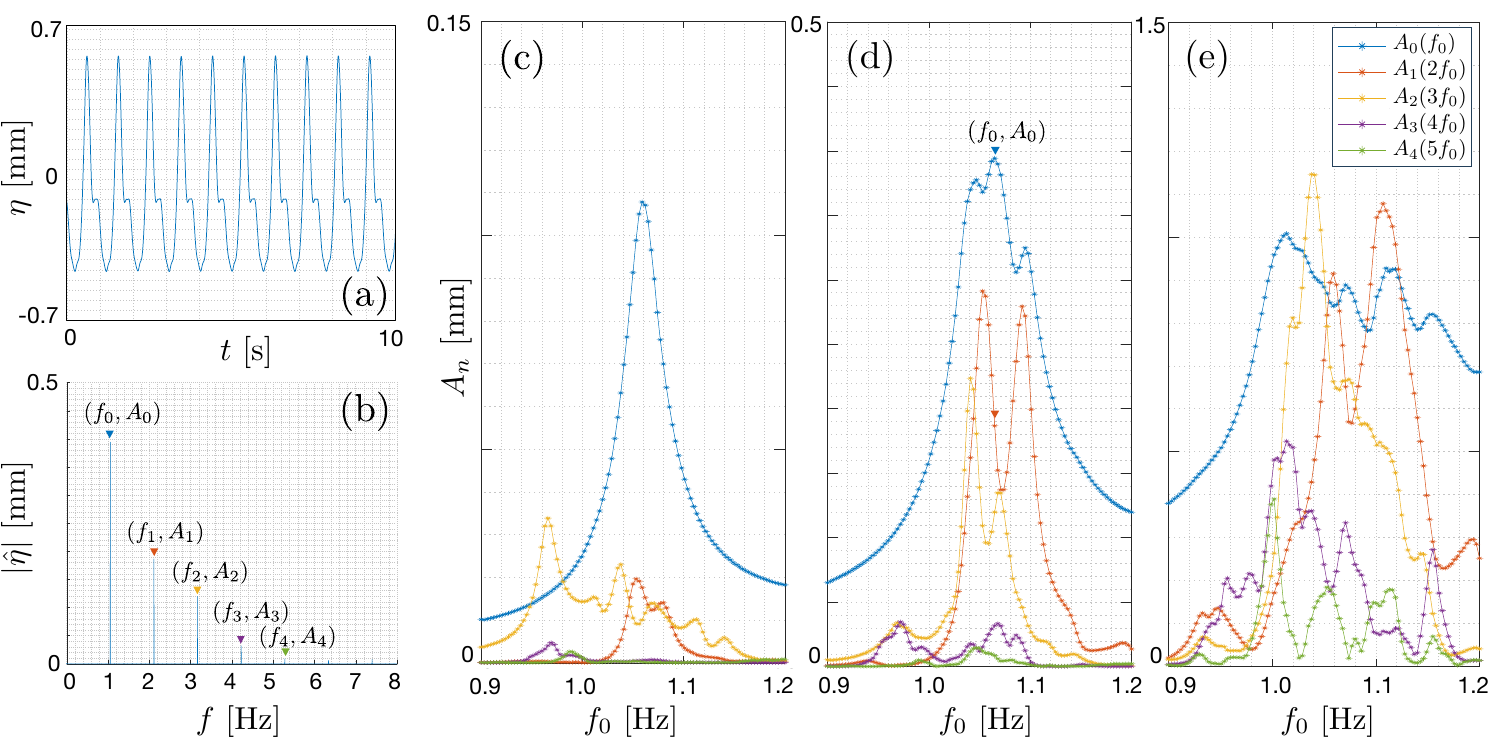}
        \caption{Example of the time signal [associated with the amplitude peak of the fundamental frequency at (b)] measured by a diplacement sensor (a) and its corresponding spectrum (b) obtained by the Fourier transform. Resonant curves of the fundamental frequency $A_0(f_s)$ (blue curve), the first harmonic $A_1(f_s)$ (orange curve), the second harmonic $A_2(f_s)$ (yellow curve), the third harmonic $A_3(f_s)$ (purple curve), and the fourth harmonic $A_4(f_s)$ (green curve) for (c) $A_s=0.7 \, \mathrm{mm}$, (d) $A_s=2.5 \, \mathrm{mm}$, and (e) $A_s=15 \, \mathrm{mm}$.}
        \label{nonlin3}
        \end{figure*}

Free-surface elevation is measured using two Keyence CL-P070 confocal laser displacement sensors. These sensors employ a multi-color confocal method: they emit several wavelengths at once, each with a distinct focal distance, and determine the distance to the surface by detecting which color is in focus on the target (free surface) \cite{keyence2025laser}. Only the wavelength that is precisely in focus at that specific distance is sharply reflected back and detected by a spectrometer, which identifies the wavelength and correlates it to the exact position of the surface. This provides high-precision ($\pm 2.2 \, \mathrm{\mu m}$), non-contact height measurements that are robust to surface reflectivity (unlike classical triangulation \cite{berkovic2012optical}). These two confocal displacement sensors (with an acquisition frequency of 1~kHz), positioned above the channel [as shown in Fig.~\ref{setup}(b)], are spaced at an interval of $10 \, \mathrm{cm}$, corresponding to the length of the segment, permitting to measure at two cross section changes simultaneously. The sensors are connected to the trolley to change its position and measure at different points along the channel (every 2.5 cm).

\subsection{Linear regime}
We report in Fig.~\ref{exp1}(a) the average of all measured spectra for the periodic channel with $w_1=5 \, \mathrm{cm}$ and $w_2=10 \, \mathrm{cm}$ ($s=1/3$ in Eq. \eqref{coupling}). The edge mode is visible at $f_E=1.097 \, \mathrm{Hz}$ and is represented by the most prominent resonance peak surrounded by the large band gap. The vertical dashed lines in the figures represent the eigenvalues from the numerical solution of the two-dimensional Helmholtz eigenproblem \eqref{helmSSHeigen}, showing excellent agreement with the experimental data. The same procedure is applied to the rectangular channel, characterized by  $s=1/2$.  In this scenario, the resonance modes are observed to be equally spaced [as shown in Fig.~\ref{exp1}(b)],  aligning with the expected regular modes of a rectangular cavity ($kL=m\pi$ with $m=1,2,\dots $). Fig.~\ref{exp1}(c) presents a comparison involving the experimental values (represented by dots), the one-dimensional predictions of the SSH model (illustrated by solid curves), and the solutions of the two-dimensional eigenproblem obtained numerically (indicated by dashed curves). The experimentally obtained local maxima for $s=1/3$ (blue dots) and $s=1/2$ (orange dots) are notably indicated. There is an excellent concordance between the theoretical predictions and experimental measurements, with minor discrepancies between 1D and 2D models attributable to near-field effects on each cross section change.

Fig.~\ref{exp2}(a) presents the profile of the absolute free surface elevation of the edge mode, with an excellent agreement obtained between the experiment and both 2D and 1D SSH prediction. Note that in our closed-cavity geometry the sloshing edge mode is localized at the left end of the cavity, with an amplitude that is negligible on the right edge. It means that  we could have replaced the rigid wall at the right end with an absorbing beach without changing the edge mode (however, this absorbing beach would influence strongly the other sloshing modes that are not localized at the edge). The collection of experimental results is showcased in Fig.~\ref{exp2}(b) and compared with the numerical simulations conducted via the Finite Element Method depicted in Fig.~\ref{exp2}(c). The simulation is carried out as follows: we solve the two-dimensional Helmholtz equation \eqref{helmSSH} with a source term $s(x,y)$, for wavenumbers $k=\kappa + \mathrm{i}\beta$, where the imaginary part of the wavenumber accounts for an adhoc viscous attenuation following the bulk law $\beta=4 \kappa^{2} \nu \omega/ g$, with $\nu$ the kinematic viscosity. The source is modelled as a Gaussian bell of the form
    \begin{align}
    s(x, y)=\frac{1}{2\pi\sigma_x\sigma_y} \exp \left[-\left(\frac{\left(x-x_{0}\right)^{2}}{2 \sigma_{x}^{2}}+\frac{\left(y-y_{0}\right)^{2}}{2 \sigma_{y}^{2}}\right)\right],
    \end{align}
where $\sigma_x=\sigma_y=w_1/30$, $x_0=d/2$, and $y_0=w_2/2$.
The profile of the wave is obtained by averaging the result along the $y$ axis and is presented in Fig.~\ref{exp2}(c) for different frequencies $f=\omega/2\pi$, where $\omega^2=gk\tanh(kh)$. The edge mode, prominently localized on the left side of the channel, is clearly discernible and aligns closely with the numerical simulation results. This mode is encapsulated within a band gap, beyond which the bulk extended modes of the channel manifest.

\subsection{Nonlinear regime}
Nonlinear effects in water waves occur readily, and in our experiment, as the source amplitude (and thus the wave amplitude) increases, secondary peaks arise around the resonance peak of the edge mode. To delve deeper into this phenomenon, we now analyze the dependence of the spectrum on amplitude variations.

Since we are no longer in the linear regime, the approach using a chirp signal that benefits from the linear properties of the Fourier transform can no longer be used. Therefore, the point source realizes vertical sinusoidal motion with the frequency $f_s$ and the amplitude $A_s$. The set of measurements, focused around the frequency of the edge mode, is studied for different values of $f_s$ and $A_s$, i.e.,  $A_{s} \in[0.5,15] \, \mathrm{mm}$, $ f_{s} \in[0.9,1.2] \, \mathrm{Hz}$, $\Delta f_{s}=0.003 \,\mathrm{Hz}$. For each source amplitude $A_s$, the amplitude of the wave is registered using a confocal displacement sensor at $x/d=4$. The signal of the length $t=20 \, \mathrm{s}$ is measured for a given frequency $f_s$, then the frequency is increased by $\Delta f_s$, and the stationary state of the wave is anticipated before registering the next signal. Usually it takes around $80 \, \mathrm{s}$ to measure one signal for the pair $(A_s,f_s)$. Due to the time consuming procedure, an automatic script is put in place that changes both $A_s$ and $f_s$ and registers the signals. To avoid the change of the water properties, the channel is covered with a transparent foil preventing the surface from being polluted, and also to suppress the evaporation that would result in decreasing the water depth and therefore changing the resonant frequencies. The signal is trimmed to obtain an integer number of periods to extract the exact values of its spectrum. The determination of the resonance curves is done by extracting the values of the fundamental frequency, first five harmonics ($f_0,\dots,f_5$) and its amplitudes ($A_0,\dots,A_5$). 

\begin{figure}[!t]
    \centering
    \includegraphics[width=0.38\textwidth]{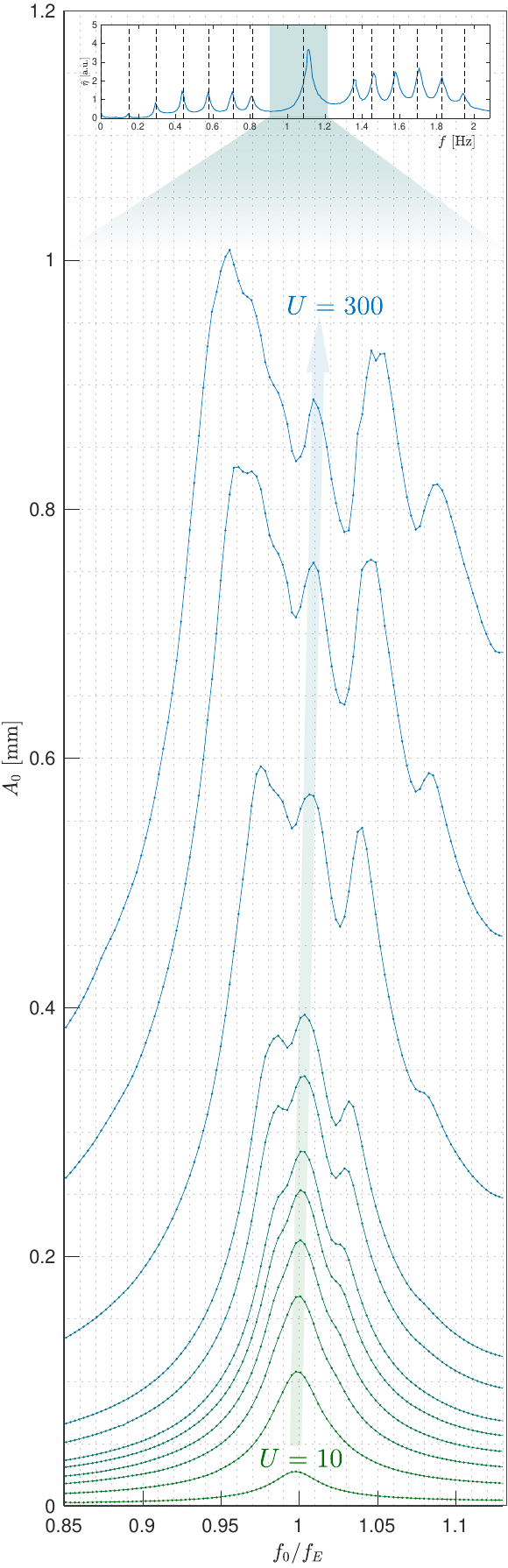}
    \caption{Resonant curves for the fundamental frequency for different Ursell numbers. The inset shows the region around the edge mode peak, where the measurements are carried out.} 
    \label{nonlin5}
    \end{figure}

One measurement for the pair of $(A_s,f_s)$ [Fig.~\ref{nonlin3}(a-b)] represents one point on the resonance curves [Fig.~\ref{nonlin3}(c-e)]. We report in Fig.~\ref{nonlin3} three different regimes of the resonant behavior of the edge mode for $A_s=0.7 \, \mathrm{mm}$ [Fig.~\ref{nonlin3}(c)], $A_s=2.5 \, \mathrm{mm}$ [Fig.~\ref{nonlin3}(d)], and $A_s=15 \, \mathrm{mm}$ [Fig.~\ref{nonlin3}(e)].  As the driving frequency $f_s$ corresponds almost exactly to the fundamental frequency $f_0$ and the harmonics ($f_1=2f_0, f_2=3f_0,\dots $) (with an error smaller than 1\%) we compare them on the same horizontal axis $f_0$. We can see that for $A_s=0.7 \, \mathrm{mm}$ we obtain only one resonant peak of the edge mode, and the contribution of higher harmonics is relatively small (less than 20 \%). For the amplitude of the source of $A_s=2.5 \, \mathrm{mm}$ [Fig.~\ref{nonlin3}(d)] secondary peaks around the main resonant peak of the edge mode appear (first bifurcation). The contribution of the harmonics becomes more important. Note that the peak on the right side of the peak of the fundamental frequency (blue curve) of the edge mode corresponds to the maximum of the second harmonic (yellow curve). On the other hand, the left side peak correlates with the maximum of the first harmonic (orange curve). With further increase of the amplitude $A_s$ we can observe the emergence of additional peaks (second bifurcation) reported in Fig.~\ref{nonlin3}(e) where the resonant curves for $A_s=15 \, \mathrm{mm}$ are shown. The amplitudes of the first and second harmonic are now higher than the fundamental frequency amplitude. Experimental measurements show the existence of nonlinear interactions near the resonant frequency of the edge mode. Note that similar behavior for the rectangular channel was obtained experimentally and described theoretically in \cite{khosropour1995resonant, chester1968resonant, chester1968resonantExp}. The apparent resonant energy transfer between the fundamental frequency and its harmonics, as described in \cite{miles1985resonantly}, can be analyzed using a modal expansion approach, where the evolution of mode amplitudes follows a set of weakly nonlinear coupled equations. To further investigate the emergence of the secondary resonances, we introduce the Ursell number as $U=A_s\lambda^2/h^3$, where $\lambda \approx 4d$ stands for the wavelength of the edge mode and $h$ denotes the water depth. In our case, the system is deeply in the shallow water regime since $\lambda=40\, \mathrm{cm}$ is much larger than the water depth $h=2\, \mathrm{cm}$ ($\tanh(kh)/(kh) \approx 0.97$). This condition allows the long-wave approximation to apply, making the Ursell number a valid descriptor for the observed phenomena. The ensemble of the measurements is shown in Fig.~\ref{nonlin5}. It appears that the first bifurcation, where two secondary peaks emerge, happens when the Ursell number $U\approx 25$. The peaks are placed almost symmetrically at around $\pm 5 \% f_E$ away from the main resonance of the edge mode ($f_E$). The second bifurcation, i.e., when additional peaks arise at approximately $\pm 10 \% f_E$ away from the original edge mode resonance, occurs at $U\approx 100$. 

\section{Conclusion}
The main objective of the presented work is to experimentally investigate the topologically protected edge states and band gaps in a water waveguide with periodic geometry which can be mapped to the Su-Schrieffer-Heeger model. A waveguide with step periodic width ($s=1/3$) is manufactured and examined using confocal displacement sensors allowing the measurement of water free surface elevation. Two-dimensional numerical simulations are carried out and compared to the discrete SSH model and experimental data. The obtained results show that this very simple setup exhibits all the properties of the SSH model with an excellent agreement to the water wave systems. Furthermore, the system is analyzed in a nonlinear regime, revealing two distinct bifurcation regimes. The first bifurcation corresponds to the emergence of two secondary resonances around the primary edge wave resonance for Ursell number $U>25$. The second bifurcation, with the appearance of additional peaks around the main edge mode resonance, is recognized for $U>100$. This phenomenon, tentatively interpreted as an energy transfer between the fundamental frequency and harmonics, warrants further theoretical investigation for comprehensive understanding.

\bibliography{apsSSH}
\end{document}